\documentclass[aps,draft,twocolumn,showpacs,
preprintnumbers,amsmath,amssymb,footinbib]{revtex4}
\newcommand{\beq}{\begin{equation}}
\newcommand{\eeq}{\end{equation}}

\def\anp#1#2#3{Annals Phys. {\bf #1}, #2 (#3)}
\def\cnpp#1#2#3{Comments Nucl. Part. Phys. {\bf #1}, #2 (#3)}
\def\ibid#1#2#3{{\it ibid.} {\bf #1}, #2 (#3)}
\def\ijm#1#2#3{Intl. Jour. Mod. Phys. {\bf #1}, #2 (#3)}
\def\npa#1#2#3{Nucl. Phys. A {\bf #1}, #2 (#3)}
\def\npb#1#2#3{Nucl. Phys. B {\bf #1}, #2 (#3)}

\def\plb#1#2#3{Phys. Lett. B {\bf #1}, #2 (#3)}
\def\prc#1#2#3{Phys. Rev. C {\bf #1}, #2 (#3)}
\def\prd#1#2#3{Phys. Rev. D {\bf #1}, #2 (#3)}
\def\prl#1#2#3{Phys. Rev. Lett. {\bf #1}, #2 (#3)}
\def\phr#1#2#3{Phys. Rep. {\bf #1}, #2 (#3)}
\def\ppnp#1#2#3{Prog. Part. Nucl. Phys. {\bf #1}, #2 (#3)}

\begin{document}
\title{Dilute Nuclear Matter in Chiral Perturbation Theory}
\author{E. S. Fraga$^{a,b}$, Y. Hatta$^{c,d}$, R. D. Pisarski$^e$,
and J. Schaffner-Bielich$^{f,g}$}
\affiliation{
$^a$Laboratoire de Physique Th\'eorique,Universit\'e Paris XI
B\^atiment 210, 91405 Orsay Cedex, France\\
$^b$Instituto de F\'\i sica,
Universidade Federal do Rio de Janeiro
C.P. 68528, Rio de Janeiro, 21941-972 RJ, Brazil\\
$^c$Department of Physics, Kyoto University, Kyoto 606-8502, Japan\\
$^d$The Institute of Physical and Chemical Research (RIKEN)
Wako, Saitama 351-0198, Japan\\
$^e$High Energy Theory, Department of Physics, Brookhaven National Laboratory
Upton, NY 11973-5000, USA\\
$^f$Department of Physics, Columbia University
538 W. 120th Street, New York, NY 10027, USA\\
$^g$Institut f\"ur Theoretische Physik, J. W. Goethe-Universit\"at
D-60054 Frankfurt/Main, Germany}
\date{\today}

\begin{abstract}
We use chiral perturbation theory to compute the effective nucleon
propagator in an expansion about low density in the chiral limit. 
We neglect four-nucleon interactions and focus on pion 
exchange. Evaluating the nucleon self-energy on its mass shell to 
leading order, we show that the effective nucleon mass increases 
by a small amount. We discuss the relevance of our results to 
the structure of compact stars.
\end{abstract}

[]
\maketitle

As nuclear matter is compressed, eventually a transition to quark-gluon
matter occurs.  By asymptotic freedom, at very high densities
the equation of state can be computed in perturbation theory
\cite{pert,hdl,previous}.  This can be extended to moderate
densities by various approximation schemes \cite{hdl,previous}.
At low densities, the conventional approach is to
use phenomenological potentials to fit observed properties of nuclear
matter \cite{akmal}, and then extrapolate up in density.

How the nuclear equation of state matches
onto that for quark matter is of great significance for 
astrophysics \cite{previous,previous_others}.  
The standard expectation, as in 
Quantum Hadrodynamics for example \cite{qhd}, is that
hadronic pressure rises quickly to a value near that for an ideal
Fermi gas of quarks and even exceeds it at densities above normal
nuclear density. In this case there is only one type
of hadronic star, which might have a (small) quark core.
If the hadronic pressure is small relative to that of ideal
quarks, though, then there can be two classes of hadronic stars.  
There are ``ordinary'' neutron stars, which are mainly
composed of nucleons.
In addition, there are stars with a large quark core;
their mass and radius are (approximately)
half that of ordinary neutron stars. In addition, the pion tadpole
contribution does not depend on the density to leading order.

Thus it is imperative to understand the equation of state
for nuclear matter. In Ref. \cite{savage+wise}, Savage and Wise 
compute mass shifts for the baryon octet using chiral perturbation 
theory, including all operators which contribute to leading order
in the density.  Due to four-nucleon interactions, they
find that all masses decrease with increasing density; extrapolating
to nuclear matter densities, the shifts are considerable.
The self-energies were computed at zero momentum, though, while
the physical point is on the mass shell.  In the presence of a 
Fermi sea, the mass shell changes, in a way which is easily
computed.  To leading order in the density, the difference
in mass shell only affects exchange, and not contact, terms.
In this paper we compute the nucleon self energy on its mass shell, from
the diagram for pion exchange.  
In this case, unlike Ref. \cite{savage+wise}, the usual
logarithms of chiral perturbation theory appear on the mass shell.
We note that the shift in the nucleon mass is dominated by
contact terms, not single pion exchange,
so the following exercise is a minor point of principle.

We assume that nucleons, of mass $m$, are heavy, and that
pions, with mass $m_\pi$, are very light. 
The interaction of pions and nucleons is determined by 
the spontaneous breaking of
chiral symmetry 
\cite{donoghue,horowitz,qhd,kaiser,meissner}.  For
light pions, to leading order in chiral perturbation theory, 
the only parameter which enters is the pion decay constant, 
$f_\pi\approx 93$~MeV.  

The proceeding calculation is elementary, and, besides those
of Savage and Wise \cite{savage+wise}, it is similar to
computations by Horowitz and Serot \cite{horowitz}, 
by Bernard, Kaiser, and Mei\ss ner \cite{kaiser}, and by
Mei\ss ner, Oller, and Wirzba \cite{meissner}.  

The parameters which enter can be understood without explicit
computation.  A Fermi gas of 
nucleons is characterized by a Fermi momentum,
$p_f$; up to and including nuclear matter densities, $p_f \ll m$.
In an expansion about low densities, the natural parameter
which enters into the nucleon propagator 
is just the density, $n_{\rm nucl}\sim p_f^3$.  We would
like a dimensionless parameter to characterize the expansion.
At one loop order, chiral perturbation theory brings in
two powers of $1/f_\pi$.  The only other parameter in the
problem is the nucleon mass, $m$ (at least for vanishing
pion mass).  Thus to leading order,
the corrections to the nucleon propagator are proportional to
\beq
\frac{p_f^3}{m f_\pi^2} ,
\label{e1}
\eeq
which we now compute.

To leading order in chiral perturbation theory, we take the
nucleon Lagrangian to be
\beq
{\cal L} =
\overline{\psi} \left(
\not \! \partial - \gamma^0 \, \mu + m -  
i \frac{g_A}{f_\pi} \not \! \partial \pi \gamma_5 \right)  
\psi \; ,
\eeq
where $\mu = \sqrt{p_f^2 + m^2}$ is the chemical potential,
$g_A \approx 1.2$ is the axial vector coupling constant,
and $\pi = \pi^a \sigma^a/2$, where
the $\sigma^a$'s are Pauli matrices in $SU(2)$ flavor.
Other interactions, such as between two nucleons and more than
two pions, involve more powers of $1/f_\pi$, and so enter beyond
leading order in the density.
There are tadpole contributions, with a pion in the loop,
but like the tadpoles from four-nucleon interactions, these
are independent of the external momentum, and so of the choice
of mass shell.

Thus we consider single pion exchange, which contributes 
to the nucleon self energy, $\Sigma$, as
$$
\Sigma(P) =
- \frac{3\, g_A^2}{4\, f_\pi^2} 
\int \frac{d^4K}{(2 \pi)^4} \; \frac{1}{(P-K)^2 + m_\pi^2} \; \times
$$
\beq
\gamma_5 (\not \! P - \not \! K) 
\frac{1}{- i \not \! K - \gamma^0 \mu + m} 
\gamma_5 (\not \! P - \not \! K) ;
\eeq
$P=(p^0,\vec{p})$ is the four-momentum of the nucleon.
The diagrams are evaluated using the imaginary time formalism.

In general, the nucleon self energy $\Sigma(P)$ is a rather 
complicated function of $p^0$ and $\vec{p}$ \cite{horowitz,kaiser}.
Here we shall only compute the nucleon self energy at a 
special point, on its mass shell:
\beq
p^0 =
p^0_{ms} = i (\mu - E_p) \approx i \, \frac{(p_f^2 - p^2)}{2 m} 
\; ;
\label{massshell}
\eeq
$E_p$ is the energy of a nucleon with momentum $p$, so $\mu = E_{p_f}$,
and $E_p \approx m + p^2/(2m)+\ldots$.
We also assume that
the momentum $p$ is on the order of the Fermi momentum, but it
need not be especially near $p_f$.  

Working on the mass shell allows us to greatly simplify the
calculation.  
As we are working at nonzero fermion density, it
is convenient to do the integral over $k^0$ first, and then
integrate over $\vec{k}$.  
In the imaginary time formalism, we
first compute the diagram for real $p^0$, and then analytically
continue to imaginary values of $p^0$, as in (\ref{massshell}).

In the integrand, there are four poles:
two from the pion propagator, 
at $k^0 = p^0 \pm i \sqrt{(\vec{k}-\vec{p})^2 + m_\pi^2}$, and two
from the nucleon propagator, at 
$k_0 = i (\mu \pm E_k)$.  
Closing the contour in imaginary $p^0$ plane, only those poles
in the upper half plane contribute.

All we are interested in, though, are the density dependent effects.
Of the four poles in the one loop diagram for the nucleon
propagator, clearly one is special.  The pole at which
$
k_0 = i \left( \mu - E_k \right) 
$
is in the upper half plane when $k < p_f$, and moves into the
lower half plane when $k > p_f$.  For the other three poles,
the sign of their imaginary part does not change with $k$.

All of the density dependent effects in the one loop diagram for
the nucleon propagator are due to the shift in this one pole.
To see this, note that the chemical potential $\mu$ only enters
by changing $ip^0 \rightarrow ip^0 + \mu$.  If we work on the
mass shell, however, then $ip^0 + \mu = + E_p$; while $p^0$
changes with $\mu$, $ip^0 + \mu$ does not.  Thus if no poles
switched the sign of their imaginary part, then we would find
that there were no density dependent effects in the propagator
at one loop order.  For example, there is wave-function
renormalization for the nucleon field, but given that $\mu$ enters
just as a shift in $p^0$, this is standard; there is no new
wave-function renormalization associated with $\mu \gamma^0$, separate
from $\not \! p$.

The contribution of the pole at 
$k^0 = i \left( \mu - E_k \right)$ is simple to
include: one only integrates over momentum below the Fermi
surface.  A similar result is found, rather more immediately,
using the real time formalism.  There, the nucleon propagator
is the sum of two terms, one the same as in the vacuum, plus
a density dependent term.  

To pick up the contribution of just this one pole, we take
\beq
\int \frac{d k^0}{2 \pi} 
\frac{1}{- i \not \! K - \gamma^0 \mu + m}
= - \frac{1}{2 E_k}
\left( i \not \! K + \gamma^0 \mu + m \right) .
\eeq
On the right hand side, $k_0 = i (\mu - E_k)$, and only
$|k| < p_f$ contribute to the integral over $\vec{k}$.

Now we need to sandwich the inverse nucleon propagator in this
expression between the $\gamma_5 (\not \! P - \not \! K)$'s from
the pion vertices.  There are three types of terms which contribute.
One is from the term $\sim m$ in the nucleon propagator:
\beq
c_m  = \gamma_5 (\not \! P - \not \! K) m  
\gamma_5 (\not \! P - \not \! K) = - m (P-K)^2 ,
\label{e2}
\eeq
one from the term $\sim \mu \gamma^0$:
\beq
c_\mu = \gamma_5 (\not \! P - \not \! K) \mu \gamma^0
\gamma_5 (\not \! P - \not \! K)
\label{e3}
\eeq
$$
= \gamma^0 \mu \left( (p^0 - k^0)^2 - (\vec{p} - \vec{k})^2 \right)
+ 2 \mu (p^0 - k^0) 
\left( \not \! \vec{p} - \not \! \vec{k} \right),
$$
and one from the term $\sim \not \! K$:
\beq
c_p \; = \; \gamma_5 \; (\not \! P - \not \! K) \; i \not \! K
\gamma_5 \; (\not \! P - \not \! K) 
\label{e4}
\eeq
$$\; = \; 
i \left( 2 (P-K)\cdot K \not \! P - (P^2 - K^2) \not \! K \right) \; .
$$

To compute the leading terms about small density, 
we can greatly simplify these expressions.  
For example, for the nucleon energy, we can replace
$E_k \approx m$, since corrections are down by $(p_f/m)^2$.  
Further, as we are computing on the mass shell,
the energy $p^0$ is small relative to the spatial
momentum; in magnitude, as $p^0 \sim p^2/m$, $p^0$ is
down by $p_f/m$ relative to $p$.  
This means that in the pion propagator, and in $c_m$, (\ref{e2}), we
can replace $(P-K)^2 \approx (\vec{p} - \vec{k})^2$.

For the other contributions, one must be careful to keep
track of relatively small terms, 
$\sim \not \! \vec{p}$, and also $\sim p^0 \gamma^0$.  
For $c_\mu$, (\ref{e3}), for the piece $\gamma^0$ we
can drop $(p^0 - k^0)^2$ relative to 
$(\vec{p} - \vec{k})^2$, and take $\mu \approx m$.
However, for the piece $\sim \mu(p^0 - k^0)$, we have
to keep track of the subdominant term, so
\beq
c_\mu \; \approx \; - \; m \gamma^0 \; (\vec{p} - \vec{k})^2
\; + \; i \; \left(p^2 - k^2\right) 
\left( \not \! \vec{p} - \not \! \vec{k} \right)\; .
\label{e6}
\eeq

For the last term, $c_p$ in (\ref{e4}), we can approximate
\beq
c_p \; \approx \; 
i \left( 2 (\vec{p}-\vec{k})\cdot \vec{k} \not \! P 
\; - \; (p^2 - k^2) \not \! K \right) \; .
\eeq
We keep the terms $\sim \gamma^0$, which are nominally down
by $p_f/m$, in order to extract the term $\sim p^0 \gamma^0$.

Adding all of these terms together, we find a remarkable
simplification:
\beq
\Sigma(p^0_{ms},p) \approx 
- \left(  \left( i \, p^0_{ms} + \mu \right) \gamma^0
+ i  \not \! \vec{p} + m \right) \Sigma_0(p) ,
\eeq
where
\beq
\Sigma_0(p) =
\frac{3 \, g_A^2}{8 m f_\pi^2}
\int_{k \leq p_f} \; \frac{d^3 k}{(2 \pi)^3} 
\left( \frac{ (\vec{p} - \vec{k})^2}{(\vec{p} - \vec{k})^2 + m_\pi^2} 
\right) \; .
\label{integral}
\eeq
We also checked that the same result is found using the real time formalism.

This form is illuminating, because it is obvious that in the chiral limit,
when $m_\pi = 0$, the function $\Sigma_0$ is independent of momentum:
\beq
\Sigma_0(p) \; = \;
+ \; \frac{ g_A^2}{16 \pi^2} \; \frac{p_f^3}{m f_\pi^2} \; = \;
+ \; \frac{3 g_A^2}{32} \; \frac{n_{\rm nucl}}{m f_\pi^2}\; , 
\label{sigma_value}
\eeq
where $n_{\rm nucl} = 2 p_f^3/(3 \pi^2)$ is the density of nucleons.  

Away from the chiral limit, $m_\pi \neq 0$, $\Sigma_0$ is
momentum dependent:
\beq
\Sigma_0(p) = 
\frac{ g_A^2}{16 \pi^2} \; \frac{p_f^3}{m f_\pi^2} 
\left( 1 + \frac{3 m_\pi^2 }{2 p_f^2} \delta \Sigma_0(p) \right) \; ,
\eeq
$$
\delta \Sigma_0(p) =
\left(\frac{p_f^2 - p^2 + m_\pi^2}{4 \, p \, p_f}\right)
\log \left( \frac{(p_f - p)^2 + m_\pi^2}{(p_f + p)^2 + m_\pi^2}\right)
$$
\beq
+ \; \frac{m_\pi}{p_f} \left( \arctan \left(\frac{p_f + p}{m_\pi}\right)
\; + \; \arctan \left( \frac{p_f - p}{m_\pi} \right) \right)
\; - \; 1 
\eeq
At zero momentum, $p=0$, this agrees with
Savage and Wise \cite{savage+wise}.  We see that
chiral logarithms appear when $p\neq0$, although
there is a $\arctan(p_f/m)$ at $p=0$.
These chiral logarithms are standard \cite{donoghue},
and relatively innocuous.
Even at the Fermi surface, $p = p_f$, they vanish like
$m_\pi^2 \log(m_\pi)$ as $m_\pi \rightarrow 0$.

One can compute $\Sigma_0$ as a function of $m_\pi$.  For illustration,
consider its value at the Fermi surface.  Then one can show
that increasing the 
pion mass tends to decrease the value of $\Sigma_0$;
as $m_\pi \rightarrow \infty$, $\Sigma_0$ vanishes like
$\approx p_f^3/(m f_\pi^2) (p_f^2/m_\pi^2)$.  
That $\Sigma_0$ vanishes like $\approx 1/m_\pi^2$ at large $m_\pi$
is evident from the integral representation, (\ref{integral}).

We can use these results to compute the nature of nucleon quasiparticles.
Adding the self energy, the effective nucleon propagator is
\beq
\Delta^{-1}_{eff}(p^0_{ms},\vec{p}) \; = \; 
\Delta^{-1}_{bare} \; - \; \Sigma
\eeq
$$
= - \left( \left( i \, p^0_{ms} + \mu \right) \gamma^0 +
i \not \! \vec{p} \right) \left( 1 - \Sigma_0 \right)
+ m \left(1 + \Sigma_0 \right) .
$$

The change in the position of the pole in the nucleon propagator is 
easy to compute.  In particular, the mass of the nucleon is shifted
{\it up}:
\beq
m_{eff} = m \left( \frac{1 + \Sigma_0}{1 - \Sigma_0} \right)
\approx m \left( 1 + 2 \Sigma_0\right) \; .
\label{e7}
\eeq
This expression holds in the chiral limit.
Comparing with (\ref{e6}), half of the mass shift arises from
the shift in the term $\sim m$, and half from what can be
viewed as wave-function renormalization.  

Away from the chiral limit, where $\Sigma_0$ is a function of
momentum, the change in the mass cannot be read off so immediately.
In that case, one has to define the effective mass by other
means, as in (11.66) of \cite{fetter}.

This increase in the effective nucleon mass is in contrast to
what happens at zero density, but nonzero temperature.
To leading order in an expansion about zero temperature,
in the chiral limit
the nucleon mass does not shift to $\sim T^2$ \cite{leutwyler}.

At normal nuclear matter density, $p_f \approx 270$~MeV.
The correction which we computed, from single pion exchange,
is tiny, $2 \Sigma_0 \approx .04$.
This suggests that chiral perturbation theory might be a reasonable
guide to the properties of nucleons, even at nuclear matter densities.

This conclusion is premature.  While the corrections to the nucleon
propagator are very small, corrections
to the pion propagator are large.
For most momentum,
such as near the pion mass shell, the corrections to the pion
propagator are like those of the nucleon, proportional to
the density, $\sim p_f^3/(m f_\pi^2)$, 
(\ref{e1}).
If the pion is far off its mass shell, though, with
an energy $\omega \sim p_f^2/m$, it can scatter
into a nucleon particle-hole pair.  
For such nearly static pions,
the pion self energy is enhanced by a factor of $m/\omega
\sim m^2/p_f^2$.  The correct expansion parameter for the
pion propagator is then not $p_f^3/(m f_\pi^2)$, but
\beq
\frac{p_f^3}{m f_\pi^2} \; \frac{m^2}{p_f^2} \; \sim \;
\left( \frac{g_A^2}{2 \pi^2} \right) \; \frac{m p_f}{f_\pi^2} \; .
\label{epion}
\eeq
Numerically, this parameter is {\it much} larger than $\Sigma_0$
in (\ref{sigma_value}).
In fact, as we are dealing with a non-relativistic
system, this enhancement of the pion propagator
is well known from condensed matter physics, and represents
the need to resum the nearly static pion propagator through
the Random Phase Approximation (RPA) 
\cite{fetter}.  Indeed, the factor of $g_A^2/(2 \pi^2)$ arises
from an explicit calculation in the RPA limit, from (4.21) of
Mei\ss ner, Oller, and Wirzba
\cite{meissner}.  For normal nuclear matter density, the parameter
of the RPA pion propagator in (\ref{epion}) is $\approx 2$ at
nuclear matter densities.  Since this parameter is only linear
in the Fermi momentum, if we require that this parameter be
less than, say, $1/2$, this means that we can use chiral
perturbation theory to compute the nuclear equation of state
only up to $p_f \sim 70$~MeV.
This corresponds to densities which are $(1/4)^3 = 1/64$ those
of normal nuclear matter!  

This restriction on the use of chiral perturbation theory is
not that surprising.  In computing the free energy, the typical
pion momentum is of order $\sim p_f$, with energies
$\sim p_f^2/m$.  To use a chiral Lagrangian, the pion momentum
should be small relative to $f_\pi$, which is
similar to the condition derived from (\ref{epion}).  What
is not evident is while there is a factor of $1/(2 \pi^2)$
from chiral perturbation theory in (\ref{epion}), this
is compensated by the factor of the nucleon mass in the numerator.

Nevertheless, such computations \cite{kaiser,meissner} are
manifestly of interest, so as to gain a more general understanding
of the nuclear equation of state.  Carrying out such calculations
beyond leading order is technically very challenging.  Using
an RPA corrected propagator for the nearly static pion is straightforward.
What is difficult is knowing how to separate diagrams with
two pion exchange from other effects.  In Quantum Hadrodynamics \cite{qhd},
one must separate two pion exchange from that of 
of heavier mesons,
such as the $\sigma$ and the $\omega$.
In ``pionless'' effective theories \cite{eff}, two pion exchange
contributes to point like interactions between four or
more nucleons.  

We can draw some tentative conclusions about the hadronic pressure,
which motivated this study.  To leading order, 
the non-ideal terms in the pressure are proportional to
$\Sigma_0$, which is very small \cite{kaiser}.
At higher order, even if corrections to the pion propagator
are large, their effect on the nucleon propagator, and the
free energy, can still be small, as a large correction to
a small number.
Thus the possibility of a hadronic phase with a small
pressure, required for a new class of quark stars, remains viable.

{\bf Acknowledgements:}
We thank E. Kolomeitsev, M. Savage, and U. van Kolck for discussions. 
E.S.F. is partially supported by CAPES, CNPq, FAPERJ and FUJB/UFRJ. 
J.S.B. is partially supported by DOE grant DE-FG-02-93ER-40764.
The research of R.D.P. is supported by DOE grant DE-AC02-98CH10886.


\begin{thebibliography}{999}
%

\bibitem{pert}
B.~A.~Freedman and L.~D.~McLerran,
Phys.\ Rev.\  {\bf D16}, 1130 (1977); 
{\it ibid.}, {\bf D16}, 1147 (1977);
{\it ibid.}, {\bf D16}, 1169 (1977);
{\it ibid.},  {\bf D17}, 1109 (1978);
V. Baluni, Phys. Rev. D {\bf 17},  2092  (1978).

\bibitem{hdl}
R.~Baier and K.~Redlich,
\prl{84}{2100}{2000}; hep-ph/9908372;
J.~P.~Blaizot, E.~Iancu and A.~Rebhan,
\prd{63}{065003}{2001}, hep-ph/0005003;
A.~Peshier, B.~K\"ampfer and G.~Soff,
\prd{66}{094003}{2002}, hep-ph/0206229.

\bibitem{previous}
E. S. Fraga, R. D. Pisarski, and J. Schaffner-Bielich,
\prd{63}{121702}{2001}, hep-ph/0101143;
\npa{702}{217}{2002}, nucl-th/0110077.

\bibitem{akmal}
A.~Akmal, V.~R.~Pandharipande and D.~G.~Ravenhall, \prc{58}{1804}{1998},
nucl-th/9804027, and references therein.

\bibitem{previous_others}
U. H. Gerlach, Phys. Rev. {\bf 172}, 1325 (1968);
N.~K.~Glendenning and C.~Kettner,
Astron.\ Astrophys.\  {\bf 353}, L9 (2000), astro-ph/9807155;
D.~Blaschke, H.~Grigorian, G.~Poghosyan, C.~D.~Roberts and S.~M.~Schmidt,
Phys.\ Lett.\  {\bf B450}, 207 (1999), nucl-th/9801060;
A.~Peshier, B.~K\"ampfer and G.~Soff,
Phys.\ Rev.\  {\bf C61}, 045203 (2000), hep-ph/9911474.

\bibitem{qhd}
B.~D.~Serot and J.~D.~Walecka,
\ijm{6}{515}{1997}, nucl-th/9701058;
R.~J.~Furnstahl and B.~D.~Serot,
\cnpp{2}{A23}{2000}, nucl-th/0005072.

\bibitem{savage+wise} 
M.~J.~Savage and M.~B.~Wise,
Phys.\ Rev.\ D {\bf 53}, 349 (1996), 
hep-ph/9507288.

\bibitem{donoghue}
J. F. Donoghue, E. Golowich, and B. R. Holstein,
``Dynamics of the Standard Model''
(Cambridge University Press, Cambridge, 1992).

\bibitem{horowitz}
C.~J.~Horowitz and B.~D.~Serot, \plb{108}{377}{1982}.

\bibitem{kaiser}
V. Bernard, N. Kaiser, Ulf-G. Mei\ss ner, 
\ijm{E4}{193}{1995}, hep-ph/9501384;
M.~Lutz, B.~Friman, and C.~Appel,
\plb{474}{7}{2000}, nucl-th/9907078;
N. Kaiser, S. Fritsch, and W. Weise,
\npa{697}{255}{2002}, nucl-th/0105057;
\npa{700}{343}{2002}, nucl-th/0108010;
\plb{545}{73}{2002}, nucl-th/0202005;
nucl-th/0212049.

\bibitem{meissner}
Ulf-G. Mei\ss ner, J. A. Oller, and A. Wirzba, 
\anp{297}{27}{2002}, nucl-th/0109026.

\bibitem{brown}
G.~E.~Brown and M.~Rho,
\prl{66}{2720}{1991};
\phr{363}{85}{2002}, hep-ph/0103102.

\bibitem{fetter}
A. L. Fetter and J. D. Walecka, 
``Quantum Theory of Many-Particle Systems''
(New York, McGraw-Hill, 1971).

\bibitem{leutwyler}
H.~Leutwyler and A.~V.~Smilga,
\npb{342}{302}{1990}.

\bibitem{eff}
S.~Weinberg, \npb{363}{3}{1991};
D.~B.~Kaplan, M.~J.~Savage and M.~B.~Wise,
\npb{478}{629}{1996}, nucl-th/9605002;
\ibid{534}{329}{1998}; nucl-th/9802075,
U.~van Kolck,
\ppnp{43}{337}{1999}, nucl-th/9902015,
J.~W.~Chen, G.~Rupak and M.~J.~Savage,
\npa{653}{386}{1999}, nucl-th/9902056,
S.~R.~Beane, P.~F.~Bedaque, M.~J.~Savage and U.~van Kolck,
\npa{700}{377}{2002}, nucl-th/0104030.
\end{thebibliography}
\end{document}